\documentclass[a4paper,11pt]{article}
\usepackage{a4wide}
\usepackage[english]{babel}  
\usepackage[utf8]{inputenc}	
\usepackage{graphicx}
\usepackage{amsmath}
\usepackage{xspace}
\usepackage{url}
\usepackage{indentfirst}
\usepackage[margin=2cm]{geometry}
\usepackage{amsfonts}
\usepackage[colorlinks=true,urlcolor=blue]{hyperref}
\usepackage{bbm}
\usepackage{amsthm}
\usepackage[all]{xy}
\usepackage{bigints}
\usepackage{color}
\usepackage{slashed}
\numberwithin{equation}{section}

\makeatletter \def\@seccntformat#1{\csname the#1\endcsname\hspace{1ex}} \makeatother

\newcommand{\D}{\mathrm{d}}

\newcommand{\LL}{\mathcal{L}}

\newcommand{\p}{\partial}

\newcommand{\ph}{\phantom}
\newcommand{\lp}{\left(}
\newcommand{\rp}{\right)}

\begin{document}
	\begin{titlepage}

		\vskip 0.4 cm
		
		\begin{center}
			{\Large{ \bf Lagrangians for non-relativistic gravity
			}}
			
			\vspace{1em}  
			
			\vspace{1em} Patrik Novosad			
			\footnote{Email address:
			  Rick.Novosad@seznam.cz  }\\
			\vspace{1em}
\textit{Department of Theoretical Physics and
				Astrophysics, Faculty of Science,\\
				Masaryk University, Kotl\'a\v{r}sk\'a 2, 611 37, Brno, Czech Republic}
			
			\vskip 0.8cm

			\vskip 0.8cm
			
		\end{center}
		
\begin{abstract}
We study the covariant expansion of Einstein-Hilbert action in powers of $1/c^2$, where $c$ is the speed of light. We assume arbitrary spacetime foliation, i.e. we separate tangent index into two groups, which depend on generic $n$. This is done firstly by suitable parametrization of geometry which is called 'pre-non-relativistic' parametrization. This allows us to rewrite the general relativity in a form suitable for the analytical $1/c^2$ expansion. 
Consequently, we can study the expansion of Einstein-Hilbert action up to the next-to-next-to-leading order.  
\end{abstract}
\end{titlepage}

\tableofcontents

\section{Introduction and Summary} 
\noindent In the recent years arose new interest in the non-relativistic theories especially the theories founded on covariant formulation of Newton gravity, known as Newton-Cartan geometry (gravity) \cite{Cartan1}, \cite{Cartan2}. There is a number of reasons for this interest, such as Quantum Hall Effect \cite{Son:2013rqa}, holography \cite{Christensen:2013lma}, \cite{Christensen:2013rfa} \cite{Harmark:2017rpg} or a possible way to the quantum gravity through the understanding of the non-relativistic string theory, for example \cite{Andringa:2012uz}-\cite{Bergshoeff:2021bmc}. (Last but not least reason to the study non-relativistic theories could be our everyday experience of only non-relativistic physics, with exception of using GPS.)

In this paper we perform expansion of Hilbert-Einstein action in parameter $c^{-2}$ with arbitrary foliation of the spacetime with respect to speed of light $c$. The first time when the covariant expansion of the GR was studied was in \cite{Dautcourt:1996pm}, more recently the study was done again with the connection to the Torsional Newton-Cartan geometry \cite{VandenBleeken:2017rij}. This work mostly follows up \cite{Hansen:2020pqs}, where the expansion was studied in very systematic way, and \cite{Hansen:2019pkl}, on which was \cite{Hansen:2020pqs} based of.    

The paper is organized in following sections: In the section \ref{expansion_sec} we present generalities of our approach to the expansion, specifically we give a form of expansion for fields with which we work and also we define a 'pre-non-relativistic' parametrization of vielbein which is suitable for the expansion. We rewrite the general relativity with usage of this parametrization and also we introduce a new covariant derivative with non-zero torsion. As a result we get Einstein-Hilbert Lagrangian which is analytical in the parameter $1/c^2$. At the end of the section we give the specific expansion of the 'temporal' and 'spatial' vielbein which play a central role in the expansion of Lagragian. In the section \ref{Lagr?sec} we start with a general expansion of Lagrangian which depend on a parameter. We expand the Lagrangian up to next-to-next-to leading order. Then we apply this expansion to Lagrangian which we got in the section \ref{expansion_sec}. With this procedure we obtain three different Lagrangians in the three different orders of expansion. We are most interested in the next-to-next-to leading order Lagrangian which we also simplify with so called 'on-shell' condition at the final part of the section.
\section{Expansion of geometry} \label{expansion_sec}
\noindent In this section we perform the expansion of the underlying geometry. We follow \cite{Hansen:2020pqs}, for another types of non-relativistic expansions, see \cite{Dautcourt:1996pm}, \cite{VandenBleeken:2017rij}. The expansion is performed in a dimensionless parameter which mimics inverse square speed of light, therefore our expansion contains only even powers of speed of light. (For expansion which also includes odd powers of speed of light see \cite{Ergen:2020yop}). Firstly we will define expansion of a generic field. To be able to use this ansantz as expansion of every field which we encounter, we have to define 'pre-non-relativistic' parametrization of a vielbein \cite{Hansen:2020pqs}. We follow with parametrization of Levi-Civita connection of General Relativity. In the zeroth order in the parametrization we find a new connection which has a non-zero torsion. This connection will be used to construct Ricci tensor which will be used in formulation of non-relativistic Lagrangian later. 
\subsection{Expansion of a field}

\noindent Our aim is to get a non-relativistic gravity from the relativistic general relativity. In other words we are interested in a expansion of general relativity around '$c = \infty$'. Because speed of light $c$ has a dimension, we need to be careful about this statement.  We rewrite speed of light to the form 
\begin{equation}
c = \frac{\hat{c}}{\sqrt{\sigma}},
\end{equation}
where $\hat{c}$ has a dimension of the speed and $\sigma$ is a dimensionless parameter in which we will expand. We also choose units in such the way that $\hat{c} = 1$. 

Our assumption is that all fields depend on speed of light and coordinates, i.e. a generic field is $\phi^{I}(\sigma, x)$, where $I$ stands for any type of indices (spacetime or internal index). We will work just with fields that are analytic in $\sigma$, i.e. they have Taylor expansion in the form 
\begin{equation}
\phi^{I}(\sigma,x) = \phi^{I}_{(0)}(x) + \sigma \phi^{I}_{(2)}(x) + \sigma^2 \phi^{I}_{(4)}(x) + \dots.
\label{TaylorGeneralExpansion}
\end{equation}
Here we make implicit statement, that we are interested just in $\frac{1}{c^2}$ expansion. If the expansion of the field does not start with $\sigma^0$, we multiply the field with a convenient factor to do. We are going to apply the ansantz for expansion to the fields in general relativity.
\subsection{Parametrization of relativistic vielbein}
\noindent Before we can use the ansatz from the previous subsection, we make so called 'pre-non-relativistic' parametrization, which is very convenient. This parametrization follows from the scaling between the time and space directions, which scale with a factor $c$ between each other. Our starting point is a relativistic vielbein $E^{A}_{\mu}$ and its inverse $E^{\mu}_{A}$ which characterize a ($d+1$)-dimensional Lorentzian manifold. The index $\mu = 0,1, \dots d$ is a spacetime index and the index $A = 0,1, \dots, d$ is a tangent space index. The key essence of 'pre-non-relativistic' parametrization is in the choosing an explicit factor of $c$ in the decomposition of vielbein. Moreover we split the tangent space index to two groups, i.e. $A=(a,a')$, where  $a=0,1,\dots, n$ and $a' = n+1, \dots, d$. The directions with unprimed index $a$ will be scaled with speed of light differently than directions with primed index $a'$. This splitting is motivated by usage of Newton-Cartan-like geometries in string theory and M-theory, for example \cite{Andringa:2012uz,Kluson:2018uss, Bergshoeff:2019pij,Kluson:2019ifd}. Therefore we write the splitting of vielbein and inverse vielbein as
\begin{align}
E_{\mu}^{A} &= c T_{\mu}^{a}\delta_{a}^{A} + \mathcal{E}_{\mu}^{a'}\delta_{a'}^{A}, \label{RelViel} \\
E_{A}^{\mu}  & =\frac{1}{c} T_{a}^{\mu} \delta_{A}^{a} +\mathcal{E}^{\mu}_{a'}\delta^{a'}_{A}.\label{RelInvViel}
\end{align}
We will called $T_{\mu}^{a}$ a 'temporal' vielbein and $\mathcal{E}_{\mu}^{a'}$ a 'spatial' vielbein. Note that fields $T_{\mu}^{a}$ and $\mathcal{E}_{\mu}^{a'}$ are still depend on $\sigma$, as all fields before expansion. We will deal with this dependence later. The tangent space indices can be risen or lowered with flat metric $\eta_{AB} = \text{diag}\,(-1,1,\dots,1)$ hence for the unprimed indices we will use metric 
\begin{equation}
\eta_{ab} = \text{diag}\,(\underbrace{-1,1,\dots,1)}_{n+1}
\end{equation}
and for the primed indices a Kronecker delta $\delta_{a'b'}$. The relativistic veilbein satisfies 
\begin{equation}
E^{\mu}_{A} E_{\nu}^{A} = \delta^{\mu}_{\nu},\quad E^{\mu}_{A} E_{\mu}^{B} = \delta_{A}^{B}.
\label{VielInvVielCons}
\end{equation}
From the (\ref{VielInvVielCons}) and from the parametrizations of vielbeins (\ref{RelViel}), (\ref{RelInvViel}) it follows that
\begin{equation}
\delta^{\mu}_{\nu} = T_{a}^{\mu}T_{\nu}^{a} + \mathcal{E}_{a'}^{\mu}\mathcal{E}_{\nu}^{a'}, \quad T_{a}^{\mu} T_{\mu}^{b} = \delta_{a}^{b},\quad T_{a}^{\mu} \mathcal{E}_{\mu}^{b'} = 0, \quad \mathcal{E}^{\mu}_{a'}T_{\mu}^{b} = 0, \quad \mathcal{E}_{a'}^{\mu} \mathcal{E}_{\mu}^{b'} = \delta_{a'}^{b'}.
\label{ProjRelViel}
\end{equation}
The relativistic vielbein transform with respect to the general coordinate transformations (GCT) generated by a vector $\Xi$ and with respect to Lorentz transformations with parameters $\Lambda^{A}_{\ph{A}B}$ as 
\begin{equation}
\delta E_{\mu}^{A} = \mathcal{L}_{\Xi} E_{\mu}^{A} + \Lambda_{\ph{A}B}^{A} E_{\mu}^{B}.
\end{equation}
The decomposition of parameter $\Lambda^{A}_{\ph{A}B}$ of Lorentz transformations has to be
\begin{equation}
\Lambda_{\ph{A}B}^{A} = \Lambda^{a}_{\ph{b}b} \delta_{a}^{A} \delta_{B}^{b}- \frac{1}{c}\Lambda^{a'}_{\ph{a}a} \delta_{a'}^{A} \delta_{B}^{a} + \frac{1}{c} \Lambda^{a}_{\ph{a}a'}\delta_{B}^{a'}\delta_{a}^{A} + \Lambda^{a'}_{\ph{a}b'}\delta_{a'}^{A} \delta_{B}^{b'}.
\label{RelLorParDecom}
\end{equation}
The factors of $c$ follow from choice which was made in (\ref{RelViel}) and (\ref{RelInvViel}). From decompositions (\ref{RelViel}), (\ref{RelInvViel}) and (\ref{RelLorParDecom}) we can conclude the transformation relations of 'temporal' and 'spatial' vielbeins to be 
\begin{align}
\delta T_{\mu}^{a} &= \mathcal{L}_{\Xi} T_{\mu}^{a} + \Lambda^{a}_{\ph{a}b} T_{\mu}^{b} + \frac{1}{c^2} \Lambda^{a}_{\ph{a}a'} \mathcal{E}^{a'}_{\mu}, \\
\delta \mathcal{E}_{\mu}^{a'} &= \mathcal{L}_{\Xi}\mathcal{E}_{\mu}^{a'} - \Lambda^{a'}_{\ph{a}a} T_{\mu}^{a} + \Lambda^{a'}_{\ph{a}b'}\mathcal{E}^{b'}_{\mu}.
\end{align}
From (\ref{VielInvVielCons}) we can deduce that the inverse vielbein transforms as 
\begin{equation}
\delta E_{A}^{\mu} = \mathcal{L}_{\Xi} E_{A}^{\mu} - \Lambda^{B}_{\ph{a}A} E_{B}^{\mu},
\end{equation}
thus after the decomposition
\begin{align}
\delta T_{a}^{\mu} & = \mathcal{L}_{\Xi}T_{a}^{\mu} - \Lambda^{b}_{\ph{b}a} T_{b}^{\mu} + \Lambda^{b'}_{\ph{a}a} \mathcal{E}^{\mu}_{b'}, \\
\delta \mathcal{E}^{\mu}_{a'} & = \mathcal{L}_{\Xi} \mathcal{E}_{a'}^{\mu} - \frac{1}{c^2} \Lambda^{b}_{\ph{a}a'} T_{b}^{\mu} - \Lambda^{b'}_{\ph{a}a'}\mathcal{E}^{\mu}_{b'}.
\end{align} 
With help of relativistic vielbein we can define metric and inverse metric 
\begin{align}
g_{\mu \nu} & := \eta_{AB}E_{\mu}^{A}E_{\nu}^{B} = c^2 T^{a}_{\mu}T^{b}_{\nu}\eta_{ab} + \mathcal{E}_{\mu}^{a'}\mathcal{E}_{\nu}^{b'}\delta_{a' b'}, \label{DefRelMetric}\\
g^{\mu\nu} & := \eta^{AB} E_{A}^{\mu}E_{B}^{\nu} = \frac{1}{c^2} \eta^{ab} T_{a}^{\mu} T_{b}^{\nu} + \delta^{a'b'} \mathcal{E}_{a'}^{\mu} \mathcal{E}^{\nu}_{b'}.
\end{align}
For convenience we define 'temporal' and 'spatial' parts of the metric 
\begin{align}
\Pi^{\mu \nu} & := \delta^{a'b'} \mathcal{E}_{a'}^{\mu} \mathcal{E}_{b'}^{\nu}, & \Pi_{\mu \nu} &:= \delta_{a'b'}\mathcal{E}^{a'}_{\mu} \mathcal{E}_{\nu}^{b'}, \nonumber \\
\mathcal{T}^{\mu \nu}& := T^{\mu}_{a} T_{b}^{\nu} \eta^{ab}, & \mathcal{T}_{\mu \nu}&:=T^{a}_{\mu} T^{b}_{\nu} \eta_{ab}.
\end{align}
They transform as
\begin{align}
\delta \Pi^{\mu \nu} & = \mathcal{L}_{\Xi} \Pi^{\mu \nu} - \frac{1}{c^2}\delta^{a'b'}\lp\Lambda_{\ph{a}a'}^{b} T_{b}^{\mu} \mathcal{E}^{\nu}_{b'} + \Lambda^{b}_{\ph{b}b'}T^{\nu}_{b}\mathcal{E}^{\mu}_{a'} \rp, \\
\delta \Pi_{\mu \nu} & = \mathcal{L}_{\Xi} \Pi_{\mu \nu} - \delta_{a'b'}\lp \Lambda^{a'}_{\ph{a}a} T^{a}_{\mu} \mathcal{E}^{b'}_{\nu} + \Lambda^{b'}_{\ph{a}a}T^{a}_{\nu} \mathcal{E}^{a'}_{\mu}\rp, \\
\delta \mathcal{T}^{\mu \nu} &= \mathcal{L}_{\Xi} \mathcal{T}^{\mu \nu} + \eta^{ab} \lp \Lambda^{b'}_{\ph{a}a} \mathcal{E}^{\mu}_{b'}T^{\nu}_{b} + \Lambda^{b'}_{\ph{b}b} \mathcal{E}_{b'}^{\nu} T_{a}^{\mu}\rp, \\
\delta \mathcal{T}_{\mu \nu}&= \mathcal{L}_{\Xi} \mathcal{T}_{\mu \nu} + \frac{1}{c^2} \eta_{ab}\lp \Lambda^{a}_{\ph{a}a'} \mathcal{E}^{a'}_{\mu} T_{\nu}^{b} + \Lambda^{b}_{\ph{a}a'}\mathcal{E}^{a'}_{\nu} T^{a}_{\mu}\rp.
\end{align}
From (\ref{ProjRelViel}) we can find, that 'temporal' and 'spatial' metric satisfy following relations 
\begin{equation}
\mathcal{T}_{\mu \nu} \Pi^{\nu \rho} = 0, \quad \mathcal{T}^{\mu \nu} \Pi_{\nu \rho} = 0, \quad \mathcal{T}_{\mu \nu} \mathcal{T}^{\nu \rho} + \Pi_{\mu \nu}\Pi^{\nu \rho} = \delta_{\mu}^{\rho}.
\end{equation}
\subsection{Parametrization of the Christoffel symbol}
\noindent The next step is an introduction of a covariant derivative. We use the fact that we defined the relativistic metric in (\ref{DefRelMetric}). From this metric we can easily construct Christoffel symbol as 
\begin{equation}
\Gamma_{\mu \nu}^{\rho} = \frac{1}{2} g^{\rho \sigma} \lp \p_{\mu} g_{\sigma \nu} + \p_{\nu} g_{\sigma \mu} - \p_{\sigma} g_{\mu \nu}\rp.
\end{equation}
We proceed further by a decomposition of this Christoffel symbol (the overscript numbers track the powers of $c^{-1}$) 
\begin{equation}
\Gamma_{\mu \nu}^{\rho} = c^2 \stackrel{(-2)}{C^{\rho}_{\mu \nu}} + \stackrel{(0)}{C^{\rho}_{\mu \nu}} + \frac{1}{c^2} \stackrel{(2)}{C^{\rho}_{\mu \nu}}, 
\end{equation}
where\footnote{We define an antisymmetrization as $B_{[\mu \nu]} : = \frac{1}{2} \lp B_{\mu \nu} -B_{\nu \mu}\rp$ and a symmetrization as $B_{(\mu \nu)}  =\frac{1}{2} \lp B_{\mu \nu} + B_{\nu \mu}\rp.$} 
\begin{align}
\stackrel{(-2)}{C^{\rho}_{\mu \nu}} & = \Pi^{\rho \lambda} \eta_{ab} \lp T_{\nu}^{a} \p_{[\mu}T_{\lambda]}^{b} + T_{\mu}^{a} \p_{[\nu}T_{\lambda]}^b\rp, \label{asdasda}\\
\stackrel{(0)}{C^{\rho}_{\mu \nu}} & = C_{\mu \nu}^{\rho} + S_{\mu \nu}^{\rho}, \\
C_{\mu \nu}^{\rho} & = T_{a}^{\rho} \p_{\mu}T_{\nu}^{a} + \frac{1}{2}\Pi^{\rho \lambda}\lp \p_{\mu} \Pi_{\lambda \nu} + \p_{\nu} \Pi_{\lambda \mu} - \p_{\lambda} \Pi_{\mu \nu}\rp, \\
S_{\mu \nu}^{\rho} &= \mathcal{T}^{\rho \lambda} \eta_{cd}\lp T_{\nu}^{d} \p_{[\mu} T_{\lambda]}^{c} + T_{\mu}^{d} \p_{[\nu}T_{\lambda]}^{c}\rp + T_{a}^{\rho}\p_{[\nu} T_{\mu]}^{a} , \\
\stackrel{(2)}{C^{\rho}_{\mu \nu}} & = - \eta^{ab} T_{a}^{\rho} \lp \Pi_{\lambda \nu} \p_{\mu} T_{b}^{\lambda} + \Pi_{\lambda \mu} \p_{\nu} T_{b}^{\lambda} + T^{\lambda}_{b} \p_{\lambda} \Pi_{\mu \nu} \rp. \label{aaaaaaaaa} 
\end{align}
We are interested in transformation with respect to GCT, because we want use any of these objects as a connection for new covariant derivative. The transformations of objects (\ref{asdasda})-(\ref{aaaaaaaaa}) with respect to GCT generated by a vector field $\Xi$ are 
\begin{align}
\delta_{\mathrm{GCT}} C^{\rho}_{\mu \nu} &= \mathcal{L}_{\Xi} C^{\rho}_{\mu \nu} + \p_{\mu} \p_{\nu} \Xi^{\rho}, \\
\delta_{GCT} S^{\rho}_{\mu \nu} & = \mathcal{L}_{\Xi}S^{\rho}_{\mu \nu}, \\
\delta_{GCT}  \stackrel{(-2)}{C^{\rho}_{\mu \nu}} &= \mathcal{L}_{\Xi}\stackrel{(-2)}{C^{\rho}_{\mu \nu}}, \\
\delta_{GCT} \stackrel{(2)}{C^{\rho}_{\mu \nu}} & = \mathcal{L}_{\Xi} \stackrel{(2)}{C^{\rho}_{\mu \nu}},
\end{align}
and we see that $C_{\mu \nu}^{\rho}$ transform as a connection. Other objects transform as tensors. With this in mind we can introduce covariant derivative as
\begin{equation}
\nabla_{\mu}A^{\nu}_{\rho} = \p_{\mu}A^{\nu}_{\rho} + C^{\nu}_{\mu \alpha} A^{\alpha}_{\rho} - C^{\alpha}_{\mu \rho} A^{\nu}_{\alpha},
\end{equation}
where $A^{\nu}_{\rho}$ is a type (1,1) tensor field. This connection has a non-zero torsion 
\begin{equation}
T_{\mu \nu}^{\rho} = 2 C^{\rho}_{[\mu \nu]} = 2 T_{a}^{\rho} \p_{[\mu} T_{\nu]}^{a}
\end{equation}
and satisfy 
\begin{align}
\nabla_{\mu} \mathcal{T}_{\nu \rho} &= 0, \\
\nabla_{\mu} \Pi^{\nu \rho} &= 0, \\
\nabla_{\mu} \mathcal{T}^{\nu \rho} &= \Pi^{\lambda(\rho} \mathcal{T}^{\nu)\alpha}\left[ \p_{\alpha} \Pi_{\lambda \mu} - \p_{\mu} \Pi_{\lambda \alpha} - \p_{\lambda} \Pi_{\mu \alpha}\right], \\
\nabla_{\mu} \Pi_{\nu \rho} &= \mathcal{T}^{\sigma \lambda} \mathcal{T}_{\lambda (\nu|} \left[ \p_{\mu} \Pi_{|\rho) \sigma} + \p_{|\rho)} \Pi_{\sigma \mu} - \p_{\sigma} \Pi_{|\rho) \mu}\right], \\
\nabla_{\mu} T_{\nu}^{a} &= 0, \\
\nabla_{\mu} \mathcal{E}^{\nu}_{a'} &= \mathcal{E}^{\lambda}_{a'}\mathcal{E}^{\nu}_{b'} \p_{[\lambda} \mathcal{E}_{\mu]}^{b'} + \Pi^{\nu \sigma} \delta_{a'b'} \p_{[ \mu} \mathcal{E}_{\sigma]}^{b'} + \Pi^{\nu \sigma} \mathcal{E}^{\lambda}_{a'} \mathcal{E}_{\mu}^{b'}\delta_{b'c'} \p_{[\lambda} \mathcal{E}_{\sigma]}^{c'}, \\ 
\nabla_{\mu} T_{a}^{\nu} &= \frac{1}{2} \Pi^{\nu \sigma} T_{a}^{\lambda} \lp \p_{\lambda} \Pi_{\sigma \mu} - \p_{\mu} \Pi_{\sigma\lambda} - \p_{\sigma} \Pi_{\mu \lambda}\rp, \\ 
\nabla_{\mu} \mathcal{E}_{\nu}^{a'} &= \p_{[\mu} \mathcal{E}_{\nu]}^{a'} - \delta^{a'c'} \delta_{b'd'} \mathcal{E}_{c'}^{\sigma} \mathcal{E}_{(\nu}^{d'}\p_{\mu)} \mathcal{E}_{\sigma}^{b'}+ \frac{1}{2} \delta^{a'c'} \mathcal{E}_{c'}^{\sigma} \p_{\sigma} \Pi_{\mu\nu}. 
\end{align}
\subsection{Parametrization of Ricci scalar}
\noindent The next step is a decomposition of Ricci tensor followed by a decomposition of Ricci scalar. We need Ricci scalar for the construction of Einstein-Hilbert action. We define Ricci tensor as 
\begin{equation}
R_{\mu \nu} = \p_{\rho} \Gamma^{\rho}_{\mu \nu} - \p_{\mu} \Gamma^{\rho}_{\rho \nu} + \Gamma^{\rho}_{\rho \lambda} \Gamma^{\lambda}_{\mu \nu} - \Gamma^{\rho}_{\mu \lambda} \Gamma^{\lambda}_{\rho \nu}. 
\end{equation}
We decompose it as 
\begin{equation}
R_{\mu \nu} = c^4 \stackrel{(-4)}{R_{\mu \nu}} + c^2 \stackrel{(-2)}{R_{\mu \nu}} + \stackrel{(0)}{R_{\mu \nu}} + c^{-2} \stackrel{(2)}{R_{\mu \nu}} + c^{-4} \stackrel{(4)}{R_{\mu \nu}},
\end{equation}
where 
\begin{subequations}
\begin{align} 
\stackrel{(-4)}{R_{\sigma \nu}}  & =  \Pi^{\mu \rho} \Pi^{\lambda \tau} \eta_{ab} \eta_{cd} T_{\nu}^{a} T_{\sigma}^{c} \p_{[\lambda} T_{\rho]}^{b} \p_{[\tau} T_{\mu]}^{d} = \Pi^{\mu \rho} \Pi^{\lambda \tau} \p_{[\lambda} \mathcal{T}_{\rho] \nu} \p_{[\tau} \mathcal{T}_{\mu] \sigma},  \\
\stackrel{(-2)}{R_{\sigma \nu}}  & = \nabla_{\mu}\stackrel{(-2)}{C^{\mu}_{\nu \sigma}} - 2C_{[\nu \mu]}^{\lambda} \stackrel{(-2)}{C_{\lambda \sigma}^{\mu}} + \stackrel{(-2)}{C^{\lambda}_{\nu \sigma}} S^{\mu}_{\mu \lambda} - \stackrel{(-2)}{C_{\nu \lambda}^{\mu}} S_{\mu \sigma}^{\lambda} - \stackrel{(-2)}{C_{\mu \sigma}^{\lambda}} S_{\nu \lambda}^{\mu},  \\
\stackrel{(0)}{R_{\sigma \nu}} &= \mathcal{R}_{\sigma \nu} + \nabla_{\mu}S_{\nu \sigma}^{\mu} - \nabla_{\nu}S_{\mu \sigma}^{\mu} + 2 C_{[\mu \nu]}^{\lambda} S_{\lambda \sigma}^{\mu} + S^{\mu}_{\mu \lambda} S^{\lambda}_{\nu \sigma} - S^{\mu}_{\nu \lambda} S^{\lambda}_{\mu \sigma} - \stackrel{(-2)}{C^{\mu}_{\nu \lambda}} \stackrel{(2)}{C^{\lambda}_{\mu \sigma}} - \stackrel{(2)}{C^{\mu}_{\nu \lambda}} \stackrel{(-2)}{C^{\lambda}_{\mu \sigma}},  \\
\stackrel{(2)}{R_{\sigma \nu}} &= \nabla_{\mu} \stackrel{(2)}{C_{\nu \sigma}^{\mu}} + 2 C_{[ \mu \nu]}^{\lambda} \stackrel{(2)}{C_{\lambda \sigma}^{\mu}} +S_{\mu \lambda}^{\mu} \stackrel{(2)}{C_{\nu \sigma}^{\lambda}} - S_{\nu \lambda}^{\mu} \stackrel{(2)}{C_{\mu \sigma}^{\lambda}} - S_{\mu \sigma}^{\lambda} \stackrel{(2)}{C_{\nu \lambda}^{\mu}},  \\
\stackrel{(4)}{R_{\sigma \nu}} & = 4 \mathcal{T}^{\mu \tau} \mathcal{T}^{\lambda \alpha} \p_{[\tau} \Pi_{\lambda] \nu} \p_{[\mu} \Pi_{\alpha] \sigma}.
\end{align} 
\end{subequations}
We denote by $\mathcal{R}_{\sigma \nu}$ the Ricci tensor corresponding to the connection $C^{\rho}_{\mu \nu}$. We want to point out the comparison with \cite{Hansen:2020pqs} where a case with $n=0$ was investigated. For that case, the term $\stackrel{(4)}{R_{\sigma \nu}}$ is zero and $\stackrel{(2)}{R_{\sigma \nu}}$ is more simple. The last object which we need is Ricci scalar which has the following decomposition
\begin{align}
R & = g^{ \sigma \nu} R_{\sigma \nu}= \lp  \frac{1}{c^2} \mathcal{T}^{\sigma \nu} + \Pi^{\sigma\nu}\rp  \lp c^4 \stackrel{(-4)}{R_{\sigma \nu}} + c^2 \stackrel{(-2)}{R_{\sigma \nu}} + \stackrel{(0)}{R_{\sigma \nu}} + c^{-2} \stackrel{(2)}{R_{\sigma \nu}} + c^{-4} \stackrel{(4)}{R_{\sigma \nu}}\rp  \nonumber \\
& = c^{4}\stackrel{(-4)}{R} + c^{2} \stackrel{(-2)}{R} + \stackrel{(0)}{R} + \frac{1}{c^2} \stackrel{(2)}{R} + \frac{1}{c^4} \stackrel{(4)}{R} + \frac{1}{c^6}\stackrel{(6)}{R},
\end{align}
where
\begin{align}
\stackrel{(-4)}{R} &= \Pi^{\sigma \nu} \stackrel{(-4)}{R_{\sigma \nu}}, & \stackrel{(-2)}{R} &= \Pi^{\sigma \nu} \stackrel{(-2)}{R_{\sigma \nu}} + \mathcal{T}^{\sigma \nu} \stackrel{(-4)}{R_{\sigma \nu}}, & \stackrel{(0)}{R} & = \Pi^{\sigma \nu} \stackrel{(0)}{R_{\sigma \nu}} + \mathcal{T}^{\sigma \nu} \stackrel{(-2)}{R_{\sigma \nu}}, \nonumber \\
 \stackrel{(2)}{R} & = \Pi^{\sigma \nu} \stackrel{(2)}{R_{\sigma \nu}}  + \mathcal{T}^{\sigma \nu} \stackrel{(0)}{R_{\sigma \nu}}, &
\stackrel{(4)}{R} & = \Pi^{\sigma \nu} \stackrel{(4)}{R_{\sigma \nu}} + \mathcal{T}^{\sigma \nu} \stackrel{(2)}{R_{\sigma \nu}}, & \stackrel{(6)}{R} & = \mathcal{T}^{\sigma \nu} \stackrel{(4)}{R_{\sigma \nu}}. 
\end{align}
After long calculation we obtain following parts of Ricci scalar
\begin{subequations}
\label{RicciScalar}
\begin{align}
\stackrel{(-4)}{R} &= 0, \\
\stackrel{(-2)}{R} & = \Pi^{\sigma \nu} \Pi^{\mu \alpha} \eta_{ac}  \p_{[\mu} T_{\nu]}^{a} \p_{[\sigma} T_{\alpha]}^{c} , \\
\stackrel{(0)}{R} & = - 2 \Pi^{\mu \alpha} \nabla_{\mu} S_{\sigma \alpha}^{\sigma} + \Pi^{\sigma \nu} \mathcal{R}_{\sigma \nu} + 2\Pi^{\sigma \nu} \p_{[\mu} T_{\nu]}^{a} \p_{[\sigma}T_{\lambda]}^{b} \lp \mathcal{T}^{\mu \lambda} \eta_{ab} + T_{a}^{\lambda} T_{b}^{\mu} + 2 T_{a}^{\mu} T_{b}^{\lambda}\rp , \\
\stackrel{(2)}{R} &  = \mathcal{T}^{\sigma \nu} \mathcal{R}_{\sigma \nu} - 2\nabla_{\mu} \lp \mathcal{T}^{\mu \sigma} S^{\alpha}_{\alpha \sigma}\rp + 4 \mathcal{T}^{\mu \rho} T_{c}^{\lambda} T_{a}^{\nu} \p_{[\lambda} T_{\rho]}^{c} \p_{[\mu}T_{\nu]}^{a} + \mathcal{T}^{\sigma \nu} \mathcal{T}^{\mu \rho} \eta_{ac} \p_{[\mu} T_{\nu]}^{a} \p_{[\sigma}T_{\rho]}^{c}  \nonumber\\
&\ph{=} +\p_{[\sigma}T_{\beta]}^{c} \left[ T_{c}^{\nu} T_{b}^{\beta} \eta^{ab} \p_{\nu}T_{a}^{\sigma} - \mathcal{T}^{\nu \beta} \p_{\nu}T_{c}^{\sigma}\right]- 2 \eta^{ab} \nabla_{\mu} \lp T_{a}^{\mu} \nabla_{\sigma} T_{b}^{\sigma}\rp + 4 C_{[\mu \lambda]}^{\mu} \eta^{ab} T_{a}^{\lambda} \nabla_{\sigma} T_{b}^{\sigma} \nonumber \\
& \ph{=} - 2 C_{[\mu \lambda]}^{\mu}\p_{\nu} T_{e}^{\nu} \eta^{eb}T_{b}^{\lambda} +C^{\mu}_{\mu \nu} \mathcal{T}^{\nu \sigma} S_{\alpha \sigma}^{\alpha}, \\
\stackrel{(4)}{R} & =  0, \\
\stackrel{(6)}{R} & =0. 
\end{align}
\end{subequations}
We want to stress out, that all fields here are analytical in $\sigma$ and we will expand them later. In fact, there is still nothing special about this Ricci tensor, it still lead to general relativity. The Ricci tensor is just written in a convenient form for our purpose. 

\subsection{Einstein-Hilbert action}
\noindent In this subsection we introduce the form of Lagrangian, equivalent to Einstein-Hilbert Lagrangian, which we later expand to the second order in the parameter $\sigma$. Ordinary Einstein-Hilbert action is given as 
\begin{equation}
S_{\text{EH}} = \frac{c^4}{16 \pi G} \int \D^{d} x\D t  R \sqrt{-g}.
\end{equation}
We need to discuss powers of $\sigma$ here, because this Lagrangian is not analytical in $\sigma$. We have already found that in our parametrization the expansion of the Ricci scalar starts with power $\sigma^{-1}$. Because of this we define Ricci scalar which is analytical in $\sigma$ as 
\begin{equation}
\bar{R} = \sigma R.
\end{equation}
It is the same case for the volume element $\sqrt{-g}$. We find out that the volume element can be written as 
\begin{equation}
\sqrt{-g} = \sigma^{-\frac{n+1}{2}} \sqrt{- \det \lp\mathcal{T_{\mu \nu}} + \Pi_{\mu \nu} \rp}.
\end{equation}
For brevity we denote volume element as 
\begin{equation}
E = \sqrt{- \det \lp \mathcal{T_{\mu \nu}} + \Pi_{\mu \nu}\rp}.
\end{equation}
Altogether the Einstein-Hilbert action has the form 
\begin{equation}
S_{\text{EH}} =  \frac{1}{16 \pi G \sigma^{3 + \frac{n+1}{2}}} \int \D^d x \D t E \bar{R}.
\end{equation}
We denote the integrand as
\begin{equation}
\bar{\mathcal{L}} =  E \bar{R}.
\end{equation}
In the expansion of Ricci scalar (\ref{RicciScalar}) there is a couple of total derivatives, therefore we can use following identity (we assume that all boundary terms are zero)
\begin{equation}
\int \D^{d+1} x E \nabla_{\mu} V^{\mu} = \int \D^{d+1}x E 2 C^{\nu}_{[\nu \mu]}V^{\mu},
\end{equation}
where $V^{\mu}$ is a vector field, to further simplify the Lagrangian. We obtain final Lagrangian which is analytical in $\sigma$ 
\begin{align}
\bar{\LL}&  = E\Bigg[ \Pi^{\sigma \nu} \Pi^{\mu \alpha} \eta_{ac} \p_{[\mu} T_{\nu ]}^{a} \p_{[\sigma} T_{\alpha]}^{c} + \sigma\lp \Pi^{\sigma \nu} \mathcal{R}_{\sigma \nu} + 2 \Pi^{\mu \alpha} \p_{[\mu} T_{\nu]}^{a} \p_{[\sigma} T_{\alpha]}^{b}\lp \mathcal{T}^{\sigma \nu} \eta_{ab} + T_{a}^{\sigma} T_{b}^{\nu} - 2 T_{a}^{\nu}T_{b}^{\sigma}\rp\rp  \nonumber \\
& \ph{=} + \sigma^2\Big( \mathcal{T}^{\sigma \nu} \mathcal{R}_{\sigma \nu} +\mathcal{T}^{\mu \rho} \mathcal{T}^{\sigma \nu} \mathcal{T}_{\alpha \beta} C^{\alpha}_{[\mu \nu]} C^{\beta}_{[\sigma \rho]} + 2 \mathcal{T}^{\nu \lambda} \nabla_{\nu} C^{\mu}_{[\mu \lambda]} + C^{\nu}_{[\sigma \beta]} T^{\beta}_{b} \eta^{ab} \nabla_{\nu} T_{a}^{\sigma}  \nonumber\\
&\ph{=}+  C^{\rho}_{[\sigma \beta]} \Pi^{\alpha \sigma} \mathcal{T}^{\nu \beta} \nabla_{\nu} \Pi_{\rho \alpha} + 2 C^{\nu}_{[\sigma \beta]} \mathcal{T}^{\gamma \beta} C^{\sigma}_{[\gamma \nu]}\Big)\Bigg]. \label{FinalRelLag}
\end{align} 
We will expand this Lagrangian to the second order in the parameter $\sigma$ in the next section. Moreover, in the special case $n=0$, this Lagrangian can be reduced to a simple one \cite{Hansen:2020pqs}
\begin{equation}
\bar{\LL}_{n=0} = E \left( - \Pi^{\sigma \nu} \Pi^{\mu \alpha }  \p_{[\mu} T_{\nu]} \p_{[\sigma} T_{\alpha]}+ \sigma \Pi^{\sigma \nu} \mathcal{R}_{\sigma \nu} - \sigma^2 T^{\sigma} T^{\nu} \mathcal{R}_{\sigma \nu} \right).
\end{equation}

\subsection{Expansion of vielbein and other fields}
\noindent As was already mentioned a few times, the fields in previous subsections still depend on parameter $\sigma$. In this subsection we address it. We expand the vielbein and another associate fields like metrics and volume element. Recall here our assumption that fields possessed the Taylor expansion (\ref{TaylorGeneralExpansion})
\begin{equation}
\phi^{I}(\sigma,x) = \phi^{I}_{(0)}{x} + \sigma \phi^{I}_{(2)}(x) + \sigma^2 \phi^{I}_{(4)}(x) + \dots.
\end{equation}
For the vielbein we make following ansantz on the expansion
\begin{align}
T_{\mu}^{a} & = \tau_{\mu}^{a} + \sigma m_{\mu}^{a}+ \sigma^2 B_{\mu}^{a} + \mathcal{O}(\sigma^3), \label{TemVielExp} \\
\mathcal{E}^{a'}_{\mu} & = e_{\mu}^{a'} + \sigma \pi_{\mu}^{a'} + \mathcal{O}(\sigma^2). \label{SpaVielExp}
\end{align}
The fields $\tau^{a}_{\mu}$ and $e^{a'}_{\mu}$ represent the leading order terms, followed by subleading terms $m_{\mu}^{a}$ and $\pi^{a'}_{\mu}$.  
We can introduce expansion of inverse vielbein 
\begin{align}
T_{a}^{\mu} & = \tau^{\mu}_{a} - \sigma \tau^{\nu}_{a}\lp \tau^{\mu}_{b} m^{b}_{\nu} + e^{\mu}_{b'}\pi^{b'}_{\nu} \rp + \mathcal{O}(\sigma^2), \\
\mathcal{E}^{\mu}_{a'} & = e^{\mu}_{a'} - \sigma e^{\nu}_{a'}\lp \tau^{\mu}_{b} m_{\nu}^{b} + e^{\mu}_{b'} \pi^{b'}_{\nu}\rp + \mathcal{O}(\sigma^2),
\end{align}
where the leading order terms satisfy these relations 
\begin{equation}
\delta^{\mu}_{\nu} = \tau^{\mu}_{a} \tau_{\nu}^{a} + e^{\mu}_{a'}e_{\nu}^{a'}, \quad \tau_{a}^{\mu} \tau_{\mu}^{b} = \delta_{a}^{b}, \quad e^{\mu}_{b'}e^{a'}_{\mu} = \delta^{a'}_{b'}, \quad \tau^{a}_{\mu} e^{\mu}_{a'} = 0 ,\quad e^{a'}_{\mu} \tau_{a}^{\mu} = 0
\label{NonRelRel}
\end{equation}
which follow from (\ref{ProjRelViel}). In the expansion of inverse vielbeins the only degrees of freedom are in the leading term, all other terms can be deduced from it by an order by order calculation.
It is also convenient to expand 'spatial' metric 
\begin{equation}
\Pi_{\mu \nu} = h_{\mu \nu} + \sigma\Phi_{\mu \nu} +\sigma^2 \psi_{\mu \nu}+ \mathcal{O}(\sigma^3),
\label{SpatMetExp}
\end{equation}
 where we find the terms in the expansion to be 
\begin{align}
h_{\mu \nu} &= e^{a'}_{\mu} \delta_{a'b'}e^{b'}_{\nu}, \\
\Phi_{\mu \nu} & = \delta_{a'b'}\lp \pi^{a'}_{\mu} e_{\nu}^{b'} + e_{\mu}^{a'} \pi^{b'}_{\nu}\rp.
\end{align}
We skip the precise form of $\psi_{\mu \nu}$ as we won't need to work with it.   
Similarly we can expand 'inverse' 'spatial' metric, as
\begin{align}
\Pi^{\mu \nu} =  h^{\mu \nu} - 2\sigma h^{\rho(\mu} \tau^{\nu)}_{a} m_{\rho}^{a} - \sigma h^{\rho \nu} h^{\mu \tau} \Phi_{\tau \rho} + \mathcal{O}(\sigma^2), 
\end{align}
where we define 
\begin{equation}
h^{\mu \nu} =e^{\mu}_{a'} \delta^{a'b'}e_{b'}^{\nu}.
\end{equation}
It is also useful to label first term in the expansion of $\mathcal{T}^{\mu \nu}$ to be 
\begin{equation}
\tau^{\mu \nu} = \tau^{\mu}_{a} \eta^{ab} \tau^{\nu}_{b}.
\end{equation}
Moreover we need to deal with volume element. We define non-relativistic volume element to be 
\begin{equation}
e = E \Big|_{\sigma = 0} = \sqrt{- \det \lp \tau_{\mu}^{a} \eta_{ab} \tau_{\nu}^{b} + h_{\mu \nu}\rp}.
\end{equation}

In the next section we vary Lagrangian with respect to $\tau_{\mu}^{a}$ and $h_{\mu \nu}$. From (\ref{NonRelRel}) it is easy to obtain variation of the $h^{\sigma \rho}$ and $\tau^{\sigma}_{b}$ with respect to fields with the opposite position of indices
\begin{align}
\delta h^{\lambda \rho} & = - h^{\nu \lambda} h^{\mu \rho} \delta h_{\nu \mu} - 2 h^{\nu(\lambda} \tau^{\rho)}_{b} \delta \tau^{b}_{\nu}, \label{var_h} \\
\delta \tau^{\nu}_{b} &= - h^{\rho \nu} \tau^{\mu}_{b} \delta h_{\mu \rho} - \tau^{\nu}_{a} \tau^{\mu}_{b} \delta \tau^{a}_{\mu} \label{var_tau}.
\end{align} 
And we also need variation of the volume element which can be obtained from the properties of a determinant 
\begin{equation}
\delta e = \frac{1}{2} e \lp 2 \tau_{a}^{\mu} \delta \tau_{\mu}^{a} + h^{\mu \nu} \delta h_{\mu \nu}\rp.
\label{var_vol_el}
\end{equation}

\section{Non-relativistic Lagrangians} \label{Lagr?sec}
\noindent In this section we expand Lagrangian, which we found in the previous section. We begin with a general analysis of the expansion of Lagrangian. Then we apply that to the Lagrangian (\ref{FinalRelLag}), which we expand into next-to-next-to-leading order (NNLO). This non-relativistic NNLO Lagrangian will be the main result of this paper.
\subsection{Expansion of Lagrangian}
\noindent  As all other fields considered here, Lagrangian has also expansion in the powers of $\sigma$. Our Lagrangian $\bar{\LL}$ is a function of $\sigma$, $T^{\mu}_{a}$, $T^{a}_{\mu}$, $\Pi_{\mu\nu}$, $\Pi^{\mu \nu}$ and spacetime derivatives. For clarity we present here the expansion of Lagrangian which depends just on one field $\phi(\sigma,x)$ which has the expansion (\ref{TaylorGeneralExpansion}). The generalization to the case of more fields is straightforward. We are interested in the expansion of the Lagrangian  around $\sigma =0$, which is precisely non relativistic limit $ c \to \infty$. We will denote the total derivative with respect to $\sigma$ by a prime and this derivative is given by chain rule as
\begin{equation}
\frac{\D}{\D \sigma} = \frac{\p}{\p \sigma} + \frac{\p \phi}{\p \sigma} \frac{\p}{\p \phi} + \frac{\p \p_{\mu}\phi}{\p \sigma} \frac{\p}{\p \p_{\mu}\phi}.
\end{equation}
The expansion of Lagrangian is then
\begin{align}
\bar{\LL}(\sigma)\Big|_{\sigma=0} & =  \bar{\LL}(0) + \bar{\LL}'(\sigma) \Big|_{\sigma = 0} \sigma+ \frac{1}{2} \bar{\LL}''(\sigma)\Big|_{\sigma = 0} \sigma^2 + \mathcal{O}(\sigma^3)  \nonumber\\
& =\bar{\LL}(0) + \left[ \frac{\p \bar{\LL}(\sigma)}{\p \sigma} \bigg|_{\sigma = 0}+ \phi_{(2)} \frac{\delta \bar{\LL}(0)}{\delta \phi_{(0)}}  \right] \sigma \nonumber \\
& \ph{=}+ \Bigg[\frac{1}{2} \frac{\p^2 \bar{\LL}(\sigma)}{\p \sigma^2}\Bigg|_{\sigma = 0} +  \phi_{(4)} \frac{\delta \bar{\LL}(0)}{\delta \phi_{(0)}} + \phi_{(2)} \frac{\delta}{\delta \phi_{(0)}} \frac{\p \bar{\LL}(\sigma)}{\p \sigma} \bigg|_{\sigma = 0}  \nonumber\\
& \ph{=} +\frac{1}{2}\lp \phi^2_{(2)} \frac{\p^{2}\bar{\LL}(0)}{\p \phi_{(0)}^{2}} + 2 \phi_{(2)} \p_{\mu} \phi_{(2)} \frac{\p^2 \bar{\LL}(0)}{\p \phi_{(0)} \p \p_{\mu} \phi_{(0)}} + \p_{\nu}\phi_{(2)} \p_{\mu}\phi_{(2)} \frac{\p^{2} \bar{\LL}(0)}{\p \p_{\mu}\phi_{(0)} \p \p_{\nu} \phi_{(0)}}\rp \Bigg] \sigma^2 .
\end{align}
We define here three non-relativistic Lagrangians
\begin{align}
\bar{\LL}_{\text{LO}} & = \bar{\mathcal{L}}(0) = \bar{\LL}(0,\phi_{(0)}, \p_{\mu} \phi_{(0)}),\\
\bar{\LL}_{\text{NLO}} & = \frac{\p \bar{\LL}(\sigma)}{\p \sigma} \bigg|_{\sigma = 0}+ \phi_{(2)} \frac{\delta \bar{\LL}(0)}{\delta \phi_{(0)}}, \\
\bar{\LL}_{\text{NNLO}} & =  \frac{1}{2} \frac{\p^2 \bar{\LL}(\sigma)}{\p \sigma^2}\Bigg|_{\sigma = 0} +  \phi_{(4)} \frac{\delta \bar{\LL}(0)}{\delta \phi_{(0)}} + \phi_{(2)} \frac{\delta}{\delta \phi_{(0)}} \frac{\p \bar{\LL}(\sigma)}{\p \sigma} \bigg|_{\sigma = 0}  \nonumber \\
& \ph{=} +\frac{1}{2}\lp \phi^2_{(2)} \frac{\p^{2}\bar{\LL}(0)}{\p \phi_{(0)}^{2}} + 2 \phi_{(2)} \p_{\mu} \phi_{(2)} \frac{\p^2 \bar{\LL}(0)}{\p \phi_{(0)} \p \p_{\mu} \phi_{(0)}} + \p_{\nu}\phi_{(2)} \p_{\mu}\phi_{(2)} \frac{\p^{2} \bar{\LL}(0)}{\p \p_{\mu}\phi_{(0)} \p \p_{\nu} \phi_{(0)}}\rp,
\end{align}
where LO stands for leading order and NLO for next-to-leading order.
It can be shown that the following identities hold
\begin{equation}
\frac{\delta \LL_{\text{NNLO}}}{\delta \phi_{(2)}} = \frac{\delta \LL_{\text{NLO}}}{\delta \phi_{(0)}},\quad \frac{\delta \LL_{\text{LO}}}{\delta \phi_{(0)}} = \frac{\delta \LL_{\text{NLO}}}{\delta \phi_{(2)}} = \frac{\delta \LL_{\text{NNLO}}}{\delta \phi_{(4)}}.
\end{equation}
These relations imply that equations of motion for lower order Lagrangian are reconstructed in higher order Lagrangians by variation with respect to the higher order fields in the expansion. These relations can be also generalized to higher order Lagrangians and fields. 

The generalization of a situation with more fields on which Lagrangian depends is straightforward, the main difference is presence of mixed derivatives in NNLO Lagrangian
\begin{align}
\bar{\LL}_{\text{LO}}& = \bar{\LL}(0,\phi^{I}_{(0)}, \p_{\mu}\phi^{J}_{(0)}), \label{LO} \\
\bar{\LL}_{\text{NLO}} & = \frac{\p \bar{\LL}(\sigma)}{\p \sigma} \bigg|_{\sigma = 0} + \phi_{(2)}^{I} \frac{\delta \bar{\LL}(0)}{\delta \phi_{(0)}^I}, \label{NLO} \\
\bar{\LL}_{\text{NNLO}} & = \frac{1}{2} \frac{\p^2 \bar{\LL}(\sigma)}{\p \sigma^2} \bigg|_{\sigma = 0} +\phi_{(4)}^{I} \frac{\delta \bar{\LL}(0)}{ \delta \phi_{(0)}^{I}}+  \phi^{I}_{(2)} \frac{\delta}{\delta \phi^{I}_{(0)}} \frac{\p \bar{\LL}(\sigma)}{\p \sigma} \bigg|_{\sigma =0}  \nonumber  \\
& \ph{=} + \frac{1}{2}\left[\phi_{(2)}^{I}  \phi_{(2)}^{J} \frac{\p^2 \bar{\LL}(0)}{\p \phi_{(0)}^{I} \p \phi_{(0)}^{J} } + 2 \phi_{(2)}^{I} \p_{\mu} \phi_{(2)}^{J} \frac{\p^2 \bar{\mathcal{L}}(0)}{\p \phi_{(0)}^{I} \p \p_{\mu} \phi_{(0)}^{J}} + \p_{\mu} \phi_{(2)}^{I} \p_{\nu} \phi_{(2)}^{J} \frac{\p^2 \bar{\LL}(0)}{\p \p_{\mu} \phi_{(0)}^{I} \p \p_{\nu} \phi_{(0)}^{J}} \right]. \label{NNLO}
\end{align}
The fields which we use follow from expansion of relativistic 'temporal' vielbein (\ref{TemVielExp})  and 'spatial' 'metric' (\ref{SpatMetExp}) and schematically we can write
\begin{equation}
\phi_{(0)}^I = \{ \tau^{a}_{\mu}, h_{\mu \nu}\}, \quad \phi_{(2)}^I = \{m_{\mu}^{a}, \Phi_{\mu \nu}\}, \quad \phi_{(4)}^{I} = \{B_{\mu}^a, \psi_{\mu \nu}\}.
\end{equation}
\subsection{Leading order Lagrangian}
\noindent We begin with LO Lagrangian (\ref{LO}) which is the cornerstone of the whole expansion. The LO Lagrangian is 
\begin{equation}
\bar{\LL}_{\text{LO}} = \bar{\LL}(0) = E \Pi^{\sigma \nu} \Pi^{\mu \alpha} \eta_{ac} \p_{[\mu} T_{\nu]}^{a} \p_{[ \sigma}T_{\alpha]}^{c}\bigg|_{\sigma = 0}  = e h^{\sigma \nu} h^{\mu \alpha} \eta_{ac} \p_{[\mu} \tau_{\nu]}^{a} \p_{[\sigma} \tau_{\alpha]}^{c}. 
\end{equation}
The variations of $\bar{\LL}_{\text{LO}}$ with respect to fields $\tau^{a}_{\mu}$ and $h_{\mu \nu}$ are
\begin{align}
\delta \bar{\LL}_{\text{LO}} & = \left[ e \tau^{\beta}_{b} h^{\sigma \nu} h^{\mu \alpha} \eta_{ac} \p_{[\mu} \tau_{\nu]}^{a} \p_{[\sigma} \tau_{\alpha]}^{c} - 4 e h^{\beta \sigma} \tau_{b}^{\nu} h^{\mu \alpha} \eta_{ac}\p_{[\mu} \tau_{\nu]}^{a} \p_{[\sigma} \tau_{\alpha]}^{c} - 2 \p_{\mu}\lp e h^{\sigma \beta} h^{\mu \alpha} \eta_{bc} \p_{[\sigma} \tau_{\alpha]}^{c}\rp\right] \delta \tau^{b}_{\beta} \nonumber \\
& \ph{=} + \left[ \frac{1}{2} e h^{\lambda \tau} h^{\sigma \nu} h^{\mu \alpha} \eta_{ac} \p_{[\mu} \tau_{\nu]}^{a} \p_{[\sigma} \tau_{\alpha]}^{c} - 2 e h^{\sigma \lambda} h^{\nu \tau} h^{\mu \alpha} \eta_{ac}\p_{[\mu} \tau_{\nu]}^{a} \p_{[\sigma} \tau_{\alpha]}^{c}\right] \delta h_{\lambda \tau}. \label{EoMLO}
\end{align}
Note that these equations of motion can be obeyed when following condition holds 
\begin{equation}
h^{\sigma \nu} h^{\mu \alpha} \p_{[\sigma} \tau_{\mu]}^{a} = 0.
\end{equation}
For case $n=0$ this means that there is a foliation of the manifold by hypersurfaces which have the constant time coordinate. The geometry which arise from the expansion with $n=0$ is called twistless torsional Newton-Cartan geometry \cite{Christensen:2013lma}. The case $n=1$ was discussed in \cite{Hartong:2021ekg}.

\subsection{Next-to-leading order Lagrangian}
\noindent For description of an expansion of $\bar{\LL}_{\text{NLO}}$ it is useful to make couple of definitions. Let's start with Ricci tensor for $\sigma=0$ which appears in the first term of (\ref{NLO}), 
\begin{equation}
\frak{R}_{\sigma \nu} = \mathcal{R}_{\sigma \nu}\big|_{\sigma = 0}. 
\end{equation}
This Ricci tensor corresponds to the connection, which arise from $C^{\rho}_{\sigma \nu}$ with $\sigma = 0$ 
\begin{equation}
\frak{C}^{\rho}_{\sigma \nu}= C^{\rho}_{\sigma \nu} \big|_{\sigma = 0} =\tau^{\rho}_{a} \p_{\sigma} \tau^{a}_{\nu} + \frac{1}{2} h^{\rho \lambda}\lp \p_{\sigma} h_{\lambda \nu} + \p_{\nu} h_{\lambda \sigma} - \p_{\lambda }h_{\sigma \nu}\rp. 
\end{equation}
We denote covariant derivative with respect to $\frak{C}$ as $\nabla^{\frak{C}}$. Due to presence of vielbein indices we also introduce for convenience a 'torsional matrix' 
\begin{equation}
A_{ b\sigma }^{a}  = 2 \tau^{\alpha}_{b} \p_{[\alpha} \tau_{\sigma]}^{a}.
\end{equation}
Note that following relation holds 
\begin{equation}
2 \tau^{\alpha}_{b} \frak{C}_{[\alpha \sigma]}^{\lambda} = \tau^{\lambda}_{c} A_{b \sigma}^{c}.
\end{equation}
The last object what we define is a generalization of an extrinsic curvature 
\begin{align}
K_{\mu \nu a} & = \frac{1}{2} \lp \tau_{a}^{\alpha} \p_{\alpha} h_{\mu \nu} + \p_{\mu} \tau^{\alpha}_{a} h_{\alpha \nu} + \p_{\nu} \tau_{a}^{\alpha} h_{\mu \alpha}\rp = \frac{1}{2} \lp \p_{\alpha}h_{\mu \nu} -  \p_{\mu}h_{\alpha \nu}  - \p_{\nu}h_{\mu \alpha}\rp \tau^{\alpha}_{a}  \nonumber\\
&= \frac{1}{2} \lp \nabla_{\alpha}^{\frak{C}} h_{\mu \nu} - \nabla_{\mu}^{\frak{C}} h_{\alpha \nu} - \nabla_{\nu}^{\frak{C}}h_{\mu \alpha}\rp \tau^{\alpha}_{a}.
\end{align}
For $\bar{\LL}_{\text{NNLO}}$ we also need variations of the above objects which are 
\begin{align}
\delta \frak{R}_{\mu \nu} & = \nabla_{\rho}^{\frak{C}} \delta \frak{C}^{\rho}_{\mu \nu} - \nabla_{\mu}^{\frak{C}} \delta \frak{C}^{\rho}_{\rho \nu} + 2 \frak{C}^{\rho}_{[\lambda \mu]} \delta \frak{C}_{\rho \nu}^{\lambda}, \\
\delta \frak{C}^{\rho}_{\sigma \nu} & = \tau^{\rho}_{b} \nabla^{\frak{C}}_{\sigma} \delta \tau_{\nu}^{b} + h^{\mu \rho}K_{\sigma \nu b} \delta \tau_{\mu}^{b} + h^{\rho \lambda} \frak{C}_{[\sigma \lambda]}^{\alpha} \delta h_{\alpha \nu} + h^{\rho \lambda} \frak{C}_{[\nu \lambda]}^{\alpha} \delta h_{\alpha \sigma} + h^{\rho \lambda} \frak{C}_{[\nu \sigma]}^{\alpha} \delta h_{\lambda \alpha} \nonumber \\
& \ph{=} + \frac{1}{2} h^{\rho \lambda}\lp \nabla^{\frak{C}}_{\sigma} \delta h_{\lambda \nu} + \nabla_{\nu}^{\frak{C}} \delta h_{\lambda \sigma} - \nabla_{\lambda}^{\frak{C}} \delta h_{\sigma \nu}\rp, \\
\delta 2 \frak{C}^{\rho}_{[\sigma \nu]} & = 2 \tau^{\rho}_{b} \nabla_{[\sigma}^{\frak{C}} \delta \tau_{\nu]}^{b} - 2 h^{\rho \lambda} \frak{C}^{\alpha}_{[\sigma \nu]} \delta h_{\lambda \alpha}, \\
\delta A^{a}_{b \sigma} & = 2 h^{\mu \alpha} \tau^{\nu}_{b} \frak{C}^{\rho}_{[\sigma \alpha]} \tau_{\rho}^{a} \delta h_{\mu \nu} - \tau^{\mu}_{b} A^{a}_{c \sigma} \delta \tau^{c}_{\mu} + 2 \tau^{\alpha}_{b} \nabla_{[\alpha} \delta \tau^{a}_{\sigma]} + 2 \tau^{\alpha}_{b} \frak{C}^{\rho}_{[\alpha \sigma]} \delta \tau^{a}_{\rho} .
\end{align}
We can proceed further with expansion of Lagrangian. We firstly focus on a part of $\bar{\LL}_{\text{NLO}}$, which doesn't follow from $\bar{\LL}_{\text{LO}}$, i.e. 
\begin{align}
\frac{\p \bar{\LL}}{\p \sigma}\bigg|_{\sigma= 0} &= E\lp \Pi^{\sigma \nu}\mathcal{R}_{\sigma \nu} + 2 \Pi^{\mu \alpha} \p_{[\mu} T_{\nu]}^{a} \p_{[\sigma}T_{\alpha]}^{b}\lp \mathcal{T}^{\sigma \nu} \eta_{ab} + T_{a}^{\sigma} T^{\nu}_{b} - 2 T_{a}^{\nu} T_{b}^{\sigma}\rp\rp \bigg|_{\sigma = 0} \nonumber \\
& = e \lp h^{\sigma \nu} \frak{R}_{\sigma \nu} + 2 h^{\mu \alpha} \p_{[\mu} \tau_{\nu]}^{a} \p_{[\sigma} \tau_{\alpha]}^{b} \lp \tau^{\sigma \nu} \eta_{ab} + \tau^{\sigma}_{a} \tau^{\nu}_{b} - 2 \tau^{\nu}_{a} \tau^{\sigma}_{b}\rp\rp   \nonumber\\
& = e \lp h^{\sigma \nu} \frak{R}_{\sigma \nu} + \frac{1}{2}h^{\mu \alpha}\lp 8\frak{C}^{\sigma}_{[\sigma \alpha]} \frak{C}^{\nu}_{[\nu \mu]} - 4 \frak{C}^{\sigma}_{[\mu \nu]} \frak{C}^{\nu}_{[\alpha \sigma]} - A^{a}_{d \mu} A^{b}_{c \alpha} \eta^{cd} \eta_{ab}\rp\rp. \label{NLOIntPart}
\end{align}
The second part of $\bar{\LL}_{\text{NLO}}$ (\ref{NLO}) follows directly from equations of motion of $\bar{\LL}_{\text{LO}}$.

We can continue with the equations of motion. For reason presented in the last subsection of this paper, we are interested just in the variation of (\ref{NLOIntPart}). For clarity we separate variations with respect to $\tau^{a}_{\mu}$ and $h_{\mu \nu}$ to two distinct parts. We label $e h^{\sigma \nu} \frak{R}_{\sigma \nu}$ as 'ordinary' part of (\ref{NLOIntPart}) as it is only term which appears in the case $n=0$. We will call the 'new' part the remaining part of (\ref{NLOIntPart}). The $\tau$ variation of ordinary part is 
\begin{align}
\delta_{\tau}\lp e h^{\sigma \nu} \frak{R}_{\sigma \nu}\rp& = e \bigg[ \tau^{\lambda}_{c} h^{\sigma \nu} \frak{R}_{\sigma \nu} - 2 \lp h^{\lambda \sigma} h^{\rho \nu} - h^{\lambda \rho} h^{\sigma \nu}\rp \lp \nabla^{\frak{C}}_{\rho} K_{\nu \sigma c} - A^{b}_{c \rho} K_{\nu \sigma b} + 2 \frak{C}^{\mu}_{[\mu \rho]} K_{\sigma \nu c}\rp  \nonumber \\
& \ph{=} + 4 h^{\lambda \sigma} h^{\alpha \rho} \frak{C}^{\nu}_{[\sigma \alpha]} \lp 2 K_{\nu \rho c} - \tau^{b}_{\nu} \tau^{\beta}_{c} K_{\beta \rho b}\rp \bigg] \delta \tau^{c}_{\lambda}, \label{OrdTau}
\end{align}
and the $h$ variation of ordinary part is 
\begin{align}
\delta_{h} \lp e h^{\mu \nu} \frak{R}_{\mu \nu}\rp  &= e \bigg[ \frac{1}{2} h^{\mu \nu} h^{\sigma \lambda} \frak{R}_{\mu \nu} - h^{\sigma \mu} h^{\lambda \nu} \frak{R}_{\mu \nu} + h^{\nu \sigma} h^{\rho \alpha}\lp 12 \frak{C}^{\mu}_{[\mu \rho]} \frak{C}^{\lambda}_{[\nu \alpha]} + 4 \frak{C}_{[\rho \nu]}^{\mu} \frak{C}_{[\mu \alpha]}^{\lambda} + 2 \nabla^{\frak{C}}_{\alpha} \frak{C}^{\lambda}_{[\rho \nu]}\rp  \nonumber \\
& \ph{=} + \lp 4 \frak{C}^{\alpha}_{[\alpha\nu]} \frak{C}^{\mu}_{[\mu \rho]} - 2 \nabla^{\frak{C}}_{\nu} \frak{C}^{\mu}_{[\mu \rho]}\rp \lp h^{\nu \sigma} h^{\rho \lambda} - h^{\lambda \sigma} h^{\rho \nu}\rp\bigg]  \delta h_{\sigma \lambda}. \label{OrdH}
\end{align} 
This can be compared with \cite{Hansen:2020pqs}, where the case $n=0$ was studied.
The $\tau$-variation of new part is
\begin{align}
\delta_{\tau} &  \lp \frac{1}{2}e h^{\mu \alpha}\lp 8 \frak{C}^{\beta}_{[\beta \alpha]} \frak{C}^{\nu}_{[\nu \mu]} - 4 \frak{C}^{\beta}_{[\mu \nu]} \frak{C}^{\nu}_{[\alpha \beta]} - A^{a}_{d \mu} A^{b}_{c \alpha} \eta^{cd} \eta_{ab}\rp\rp  \nonumber\\
& = e \bigg[ - 4 \frak{C}^{\rho}_{[\rho \mu]} \frak{C}^{\beta}_{[\beta \alpha]} h^{\mu \alpha}\tau^{\lambda}_{e} +  \lp h^{\lambda \mu} \tau^{\alpha}_{e} - \frac{1}{2}h^{\mu \alpha} \tau^{\lambda}_{e}\rp \lp 4\frak{C}^{\beta}_{[\mu \nu]} \frak{C}^{\nu}_{[\alpha \beta]} + A^{a}_{d\mu} A^{b}_{c\alpha} \eta^{cd} \eta_{ab}\rp  \nonumber \\
& \ph{=} + 4 \lp h^{\mu \alpha} \tau^{\lambda}_{e} - h^{\lambda \alpha} \tau^{\mu}_{e}\rp \nabla^{\frak{C}}_{\mu} \frak{C}^{\beta}_{[\beta \alpha]}+\lp h^{\nu \alpha} h^{\lambda \rho} - h^{\lambda \alpha}h^{\nu \rho}\rp \lp 4 \frak{C}^{\beta}_{[\beta \alpha]} K_{\nu \rho e} - K_{\rho \nu d} A^{b}_{c \alpha} \eta^{cd} \eta_{eb}\rp \nonumber \\
& \ph{=} +  \lp h^{\rho \alpha}\tau^{\lambda}_{d} - h^{\lambda \alpha} \tau^{\rho}_{d}\rp \lp2 A^{d}_{e \alpha} \frak{C}^{\mu}_{[\mu \rho]} + 2\frak{C}^{\beta}_{[\beta \rho]} A^{b}_{c \alpha}\eta^{cd} \eta_{eb} - \nabla^{\frak{C}}_{\rho}A^{b}_{c \alpha} \eta^{cd}\eta_{eb}\rp  \nonumber \\
&\ph{=} - 2 h^{\beta \rho}h^{\lambda \alpha} \frak{C}^{\mu}_{[\alpha \beta]} K_{\mu \rho e }  + 2 \tau^{\beta}_{e} \lp h^{\nu \alpha} \nabla^{\frak{C}}_{\nu} \frak{C}^{\lambda}_{[\alpha \beta]} - h^{\lambda \alpha} \nabla_{\nu}^{\frak{C}} \frak{C}^{\nu}_{[\alpha \beta]}\rp  \nonumber \\
&\ph{=} +  h^{\mu \alpha} \tau^{\lambda}_{c} \lp  A^{a}_{d \mu} \eta^{cd} \eta_{ab} A^{b}_{e \alpha} - A^{c}_{d\mu} A^{b}_{a \alpha} \eta^{ad} \eta_{eb}\rp\bigg] \delta \tau^{e}_{\lambda}, \label{NewTau}
\end{align}
and finally the $h$-variation of new part is
\begin{align}
\delta_{h} & \lp \frac{1}{2}e h^{\mu \alpha}\lp 8 \frak{C}^{\beta}_{[\beta \alpha]} \frak{C}^{\nu}_{[\nu \mu]} - 4 \frak{C}^{\beta}_{[\mu \nu]} \frak{C}^{\nu}_{[\alpha \beta]} - A^{a}_{d \mu} A^{b}_{c \alpha} \eta^{cd} \eta_{ab}\rp\rp  \nonumber \\ 
& =   e\bigg[ \frac{1}{4} \lp h^{\mu \alpha} h^{\sigma \lambda} - 2 h^{\mu \sigma} h^{\alpha \lambda}\rp \lp 8 \frak{C}^{\beta}_{[\beta \alpha]} \frak{C}^{\nu}_{[\nu \mu]} - 4 \frak{C}^{\beta}_{[\mu \nu]} \frak{C}^{\nu}_{[\alpha \beta]} - A^{a}_{d \mu} A^{b}_{c \alpha} \eta^{cd} \eta_{ab} \rp - 8  h^{\mu \alpha}h^{\beta \lambda} \frak{C}^{\sigma}_{[\beta \alpha]} \frak{C}^{\nu}_{[\nu \mu]}  \nonumber \\
& \ph{=} + 4 h^{\mu \alpha} h^{\beta \lambda} \frak{C}^{\sigma}_{[\mu \nu]} \frak{C}^{\nu}_{[\alpha \beta]} - 2 h^{\mu \alpha} h^{\sigma \beta} \frak{C}^{\rho}_{[\mu \beta]} \tau^{a}_{\rho} \tau^{\lambda}_{d} A^{b}_{c \alpha} \eta^{cd} \eta_{ab}\bigg]  \delta h_{\sigma \lambda}. \label{NewH}
\end{align}
\subsection{Next-to-next-to-leading order Lagrangian}
\noindent In this subsection we complete the expansion of the Lagrangian (\ref{FinalRelLag}) to the second order in $\sigma$. From (\ref{NNLO}) we obtain the general form of $\bar{\LL}_{\text{NNLO}}$ as 
\begin{align}
\bar{\LL}_{\text{NNLO}} & = \frac{1}{2} \frac{\p^2 \bar{\LL}(\sigma)}{\p \sigma^2}\bigg|_{\sigma=0} + B_{\mu}^{a} \frac{\delta \bar{\LL}_{\text{LO}}}{\delta \tau^{a}_{\mu}} + \psi_{\mu \nu} \frac{\delta \bar{\LL}_{\text{LO}}}{\delta h_{\mu \nu}} + m^{a}_{\mu} \frac{\delta}{\delta \tau^{a}_{\mu}} \frac{\p \bar{\LL}(\sigma)}{\p \sigma}\bigg|_{\sigma = 0} + \Phi_{\mu \nu} \frac{\delta}{\delta h_{\mu \nu}} \frac{\p \bar{\LL}(\sigma)}{\p \sigma}\bigg|_{\sigma= 0} \nonumber \\
& \ph{=} +   \Bigg[\frac{1}{2} m_{\alpha}^{a} m_{\beta}^{b} \frac{\p^2 \LL_{\text{LO}}}{\p \tau_{\alpha}^{a} \p \tau_{\beta}^{b}} + \frac{1}{2}\Phi_{\alpha \beta}  \Phi_{\gamma \delta} \frac{\p^2 \LL_{\text{LO}}}{\p h_{\alpha \beta} \p h_{\gamma \delta}} + \Phi_{\alpha \beta} m_{\gamma}^{a} \frac{\p^2 \LL_{\text{LO}}}{\p h_{\alpha \beta} \p \tau_{\gamma}^{a}} + m_{\alpha}^{a} \p_{\beta} m_{\gamma}^{b} \frac{\p^2 \LL_{\text{LO}}}{\p \tau^{a}_{\alpha}\p \p_{\beta} \tau^{b}_{\gamma}}  \nonumber \\
& \ph{=} + \Phi_{\alpha \beta} \p_{\gamma} m_{\delta}^{a} \frac{\p^2 \LL_{\text{LO}}}{\p h_{\alpha \beta} \p \p_{\gamma} \tau_{\delta}^{a}} +  \frac{1}{2}\p_{\alpha} m_{\beta}^{a} \p_{\gamma} m_{\delta}^{b} \frac{\p^2 \LL_{\text{LO}}}{\p \p_{\alpha} \tau_{\beta}^{a} \p \p_{\gamma} \tau_{\delta}^{b}} \Bigg]. \label{GenNNLO}
\end{align}
Let us begin with the first term in (\ref{GenNNLO}) which follow easily from  (\ref{FinalRelLag})
\begin{align}\frac{1}{2} \frac{\p^2 \bar{\LL}(\sigma)}{\p \sigma^{2}}\bigg|_{\sigma=0} &= e \bigg(  \tau^{\sigma \nu} \frak{R}_{\sigma \nu} +\tau^{\mu \rho} \tau^{\sigma \nu} \tau_{\alpha \beta} \frak{C}^{\alpha}_{[\mu \nu]} \frak{C}^{\beta}_{[\sigma \rho]} + 2 \tau^{\nu \lambda} \nabla^{\frak{C}}_{\nu} \frak{C}^{\mu}_{[\mu \lambda]} + \frak{C}^{\nu}_{[\sigma \beta]} \tau^{\beta}_{b} \eta^{ab} \nabla^{\frak{C}}_{\nu} \tau_{a}^{\sigma}  \nonumber \\
&\ph{=}+  \frak{C}^{\rho}_{[\sigma \beta]} h^{\alpha \sigma} \tau^{\nu \beta} \nabla^{\frak{C}}_{\nu} h_{\rho \alpha} + 2 \frak{C}^{\nu}_{[\sigma \beta]} \tau^{\gamma \beta} \frak{C}^{\sigma}_{[\gamma \nu]} \bigg). \label{AHAHAHA}
\end{align}
We can write down contraction of Ricci tensor with 'temporal' metric
\begin{equation}
\tau^{\mu \nu} \frak{R}_{\mu \nu} = \eta^{ab}  \left[ -h^{\mu \lambda} K_{\rho \lambda a} K_{\mu \alpha b} h^{\rho \alpha} -\nabla^{\frak{C}}_{\rho}\lp h^{\mu \lambda} K_{\mu \lambda a } \tau^{\rho}_{b}\rp +  h^{\mu \nu} K_{\mu \nu a} h^{\rho \lambda} K_{\rho \lambda b} - 2 \frak{C}_{[\rho \lambda]}^{\mu} \nabla^{\frak{C}}_{\mu} \tau^{\lambda}_{a} \tau^{\rho}_{b}\right],
\end{equation}
which add up with other terms in (\ref{AHAHAHA}) to obtain
\begin{align}
\frac{1}{2} \frac{\p^2 \bar{\mathcal{L}}(\sigma)}{\p \sigma^2} \bigg|_{\sigma = 0} & = e\eta^{ab} \left[ h^{\mu \nu} h^{\rho \lambda}K_{\mu \nu a}K_{\rho \lambda b} - h^{\mu \lambda}h^{\rho \alpha} K_{\rho \lambda a} K_{\mu \alpha b} - \nabla^{\frak{C}}_{\rho} \lp h^{\mu \lambda} K_{\mu \lambda a} \tau^{\rho}_{b}\rp - 3 \frak{C}^{\mu}_{[\rho \lambda]} \nabla^{\frak{C}}_{\mu} \tau^{\lambda}_{a} \tau^{\rho}_{b}\right]   \nonumber \\
& \ph{=}+e \lp \tau^{\mu \rho} \tau^{\sigma \nu} \tau_{\alpha \beta} \frak{C}^{\alpha}_{[\mu \nu]} \frak{C}^{\beta}_{[\sigma \rho]} + 2 \tau^{\nu \lambda} \nabla_{\nu}^{\frak{C}} \frak{C}_{[\mu \lambda]}^{\mu} + \frak{C}^{\rho}_{[\sigma \beta]} h^{\alpha \sigma} \tau^{\nu \beta} \nabla_{\nu}^{\frak{C}} h_{\rho \alpha} + 2 \frak{C}^{\nu}_{[\sigma \beta]} \tau^{\gamma \beta} \frak{C}^{\sigma}_{[\gamma \nu]}\rp.
\end{align}
The second and third term in (\ref{GenNNLO}) are equations of motion of $\bar{\LL}_{\text{LO}}$ which we already obtained in (\ref{EoMLO}). The forth term in (\ref{GenNNLO}) is a variation of the part of $\bar{\LL}_{\text{NLO}}$ which we encountered in (\ref{OrdTau}) and (\ref{OrdH}). The same is true for fifth term in (\ref{GenNNLO}) which is calculated in (\ref{NewTau}) and (\ref{NewH}). Now we focus on the square bracket in (\ref{GenNNLO}) which contains the remaining terms. We denote the whole bracket with abbreviation $\big[ \dots \big]$. The particular terms in this bracket are
\begin{align*}
\frac{1}{2} m_{\alpha}^{a} m_{\beta}^{b} \frac{\p^2  \bar{\LL}_{\text{LO}}}{\p \tau^{a}_{\alpha} \p \tau_{\beta}^{b}} & = \frac{1}{2} e m_{\alpha}^{a} m_{\beta}^{b} \eta_{cd} \bigg[ \tau^{\alpha}_{a} \tau^{\beta}_{b} h^{\sigma \nu} h^{\mu \rho} - \tau^{\beta}_{a} \tau^{\alpha}_{b} h^{\sigma \nu} h^{\mu \rho} - 8\tau_{a}^{\sigma} \tau_{b}^{\beta} h^{\alpha \nu} h^{\mu \rho } + 8 \tau^{\beta}_{a} \tau^{\nu}_{b} h^{\alpha \sigma} h^{\mu \rho}  \nonumber \\
&\ph{=}+ 4 \tau^{\sigma}_{a} \tau^{\nu}_{b} h^{\alpha \beta} h^{\mu \rho} + 4 \tau^{\nu}_{b} \tau^{\rho}_{a}h^{\beta \sigma} h^{\alpha \mu} + 4 \tau^{\nu}_{b} \tau^{\mu}_{a} h^{\beta \sigma} h^{\alpha \rho}\bigg] \p_{[\mu} \tau_{\nu]}^{c} \p_{[\sigma} \tau_{\rho]}^{d} , \nonumber\\
\frac{1}{2} \Phi_{\alpha \beta} \Phi_{\gamma \delta} \frac{\p^2 \bar{\LL}_{\text{LO}}}{\p h_{\alpha \beta} \p h_{\gamma \delta}} & = \frac{1}{2} e \Phi_{\alpha \beta} \Phi_{\gamma \delta} \bigg[ \frac{1}{4}h^{\alpha \beta} h^{\gamma \delta} h^{\sigma \nu} h^{\mu \rho} - \frac{1}{2} h^{\alpha \gamma} h^{\beta \delta}h^{\sigma \nu} h^{\mu \rho} - 2 h^{\gamma \delta} h^{\alpha \sigma} h^{\beta \nu} h^{\mu \rho}  \nonumber \\
&\ph{=} + 2 h^{\alpha \sigma} h^{\beta \nu} h^{\gamma\mu} h^{\delta \rho} + 4 h^{\nu \sigma} h^{\delta \rho} h^{\alpha \gamma} h^{\beta \mu}  \bigg] \eta_{cd} \p_{[\mu} \tau_{\nu]}^{c} \p_{[\sigma} \tau_{\rho]}^{d}, \nonumber\\
\Phi_{\alpha \beta} m_{\gamma}^{a} \frac{\p^2 \bar{\LL}_{\text{LO}}}{\p h_{\alpha \beta} \p \tau_{\gamma}^{a}} & = e \Phi_{\alpha \beta} m_{\gamma}^{a} \eta_{cd} \bigg[ \frac{1}{2} \tau^{\gamma}_{a} h^{\alpha \beta} h^{\sigma \nu} h^{\mu \rho}  - \tau^{\alpha}_{a} h^{\beta \gamma} h^{\sigma \nu} h^{\mu \rho} - 2 \tau_{a}^{\gamma} h^{\alpha \sigma} h^{\beta \nu} h^{\mu \rho} - 2 \tau^{\nu}_{a} h^{\alpha \beta} h^{\gamma \sigma} h^{\mu \rho}  \nonumber \\
&\ph{=}+ 4 \tau^{\nu}_{a} h^{\alpha \gamma} h^{\beta \sigma} h^{\mu \rho} + 4 \tau_{a}^{\alpha} h^{\gamma \sigma} h^{\beta \nu} h^{\mu \rho} + 4 \tau^{\nu}_{a} h^{\gamma \sigma} h^{\alpha \mu} h^{\beta \rho}\bigg] \p_{[\mu} \tau_{\nu]}^{c} \p_{[\sigma} \tau_{\rho]}^{d}, \nonumber \\
\end{align*}
\begin{align}
m_{\alpha}^{a} \p_{\beta} m_{\gamma}^{b} \frac{\p^2 \bar{\LL}_{\text{LO}}}{\p \tau_{\alpha}^{a} \p \p_{\beta} \tau_{\gamma}^{b}} & =  2 e m_{\alpha}^{a} \p_{[ \beta} m_{\gamma]}^{b} \eta_{bc} \left[ \tau^{\alpha}_{a} h^{\mu \gamma} h^{\beta \nu} -2 \tau^{\gamma}_{a} h^{\alpha \mu} h^{\beta \nu} - 2 \tau^{\mu}_{a} h^{\alpha \gamma} h^{\beta \nu}\right] \p_{[\mu} \tau_{\nu]}^{c},\nonumber \\
\Phi_{\alpha \beta} \p_{\gamma} m_{\delta}^{a} \frac{\p^2 \bar{\LL}_{\text{LO}}}{\p h_{\alpha \beta} \p \p_{\gamma} \tau_{\delta}^{a}} & = \Phi_{\alpha \beta} \p_{[\gamma} m_{\delta]}\eta_{ac} e \lp h^{\alpha \beta} h^{\sigma \delta} h^{\gamma \rho} - 4 h^{\beta \rho} h^{\alpha \gamma} h^{\delta \sigma}\rp \p_{[\sigma} \tau_{\rho]}^{c},\nonumber \\ 
\frac{1}{2} \p_{\alpha} m_{\beta}^{a} \p_{\gamma} m_{\delta}^{b} \frac{\p^2 \bar{\LL}_{\text{LO}}}{\p \p_{\alpha} \tau_{\beta}^{a} \p \p_{\gamma} \tau_{\delta}^{b}} & =e \p_{[\alpha} m_{\beta]}^{a} \p_{[\gamma} m_{\delta]}^{b} \eta_{ab} h^{\alpha \delta} h^{\gamma \beta}.
\end{align}
We introduce the following object 
\begin{align}
F_{\mu \nu}^{c} & = \p_{\mu} m_{\nu}^{c} - \p_{\nu}m_{\mu}^{c} + 2 m_{\nu}^{a} \tau_{a}^{\rho} \p_{[\rho} \tau_{\mu]}^{c} - 2 m_{\mu}^{a} \tau^{\rho}_{a} \p_{[\rho} \tau_{\nu]}^{c} \nonumber\\
& = 2 \p_{[\mu} m_{\nu]}^{c} + 2 A_{a[\mu}^{c} m_{\nu]}^{a}.
\end{align}
For the case $n=0$, this can be thought of as a field strength for the field $m_{\mu}$. 

Because of introduction of $F^{c}_{\mu\nu}$ we can rewrite $\big[\dots\big]$ as
\begin{equation}
\big[ \dots\big] = - \frac{1}{4} e \eta_{cd} h^{\mu \rho} h^{\nu \sigma} F_{\mu \nu}^{c} F^{d}_{\rho \sigma} + \p_{[\sigma} \tau_{\rho]}^{c} h^{\sigma \delta} h^{\gamma \rho} X_{\gamma \delta c},
\end{equation}
where $X_{\gamma \delta c}$ is a tensor whose explicit form will not be need.

\subsection{On-shell condition}
\noindent In this section we discuss the simplification of $\bar{\LL}_{\text{NNLO}}$. Let us begin with recalling of equations of motion for $\bar{\LL}_{\text{LO}}$ 
\begin{align}
\tau^{b}_{\beta} & \colon e \tau^{\beta}_{b} h^{\sigma \nu} h^{\mu \alpha} \eta_{ac} \p_{[\mu} \tau_{\nu]}^{a} \p_{[\sigma} \tau_{\alpha]}^{c} - 4 e h^{\beta \sigma} \tau_{b}^{\nu} h^{\mu \alpha} \eta_{ac} \p_{[\mu} \tau_{\nu]}^{a} \p_{[\sigma} \tau_{\alpha]}^{c} - 2 \p_{\mu} \lp e h^{\sigma \beta} h^{\mu \alpha} \eta_{bc} \p_{[\sigma} \tau_{\alpha]}^{c}\rp = 0, \\
h_{\lambda \tau} & \colon e h^{\lambda \tau} h^{\sigma \nu} h^{\mu \alpha} \eta_{ac} \p_{[\mu} \tau_{\nu]}^{a} \p_{[\sigma} \tau_{\alpha]}^{c} - 4 e h^{\sigma\lambda} h^{\nu \tau} h^{\mu \alpha} \eta_{ac} \p_{[\mu} \tau_{\nu]}^{a} \p_{[\sigma} \tau_{\alpha]}^{c} = 0.
\end{align}
Both equations can be satisfied if the fields obey the condition
\begin{equation}
h^{\sigma \nu} h^{\mu \alpha} \p_{[\mu} \tau_{\nu]}^{a} = 0.
\label{onshell}
\end{equation}
This condition is of course on-shell condition. If we could apply this condition off-shell it would greatly simplify our results. From $\bar{\LL}_{\text{NNLO}}$ we can obtain equations of motion for NNLO fields $B_{\mu}^{a}$ and $\psi_{\mu \nu}$, which are the same as equations of motion from $\bar{\LL}_{\text{LO}}$ above. Moreover a lot of the terms in $\big[ \dots \big]$ has also a form of (\ref{onshell}) times some tensor. Those are reasons for the following. We variate the on-shell condition (\ref{onshell}) times an arbitrary tensor, i.e.
\begin{equation}
\delta(e h^{\mu \rho} h^{\nu \sigma} \p_{[\mu} \tau_{\nu]}^{a} X_{\rho \sigma a}),
\label{zebytobylkonec}
\end{equation}
and analyze what restrict us to apply this condition of-shell. The variation is
\begin{equation}
\delta \lp e h^{\mu \rho} h^{\nu \sigma} \p_{[\mu} \tau_{\nu]}^{a} X_{\rho \sigma a}\rp =  - e h^{\alpha \rho} \tau^{\mu}_{b} h^{\nu \sigma} \p_{[\mu} \tau_{\nu]}^{a} X_{\rho \sigma a} \delta \tau_{\alpha}^{b} - e h^{\alpha \sigma} \tau^{\nu}_{b} h^{\mu \beta} \p_{[\mu} \tau_{\nu]}^{a} X_{\rho \sigma a} \delta \tau_{\alpha}^{b} + e h^{\mu \rho} h^{\nu \sigma} \p_{[\mu} \delta \tau_{\nu]}^{a} X_{\rho \sigma a}.
\end{equation}
We see that there are few terms that spoil the possibility to apply the condition directly on Lagrangian. On the other hand, we can restrict ourselves only to some special type of variation. Particularly if we consider only a variation of the form 
\begin{equation}
\delta \tau_{\alpha}^{b} = \Omega \tau_{\alpha}^{b},
\end{equation}
where $\Omega$ is an arbitrary function of spacetime coordinates, we find out that whole variation of (\ref{zebytobylkonec}) is
\begin{equation}
\delta \lp e h^{\mu \rho} h^{\nu \sigma} \p_{[\mu} \tau_{\nu]}^{a} X_{\rho \sigma a}\rp = 0,
\end{equation}
and we can apply the on-shell condition (\ref{onshell}) directly in NNLO Lagrangian. After application of on-shell condition the particular terms in (\ref{GenNNLO}) are 
\begin{subequations}
\begin{align}
\frac{1}{2} \frac{\p^2 \bar{\mathcal{L}}(\sigma)}{\p \sigma^2} \bigg|_{\sigma = 0} & = e \eta^{ab} \left[ h^{\mu \nu} h^{\rho \lambda}K_{\mu \nu a}K_{\rho \lambda b} - h^{\mu \lambda}h^{\rho \alpha} K_{\rho \lambda a} K_{\mu \alpha b} - \nabla^{\frak{C}}_{\rho} \lp h^{\mu \lambda} K_{\mu \lambda a} \tau^{\rho}_{b}\rp - 3 \frak{C}^{\mu}_{[\rho \lambda]} \nabla^{\frak{C}}_{\mu} \tau^{\lambda}_{a} \tau^{\rho}_{b}\right]  \nonumber \\
& \ph{=}+ e \lp \tau^{\mu \rho} \tau^{\sigma \nu} \tau_{\alpha \beta} \frak{C}^{\alpha}_{[\mu \nu]} \frak{C}^{\beta}_{[\sigma \rho]} + 2 \tau^{\nu \lambda} \nabla_{\nu}^{\frak{C}} \frak{C}_{[\mu \lambda]}^{\mu} + \frak{C}^{\rho}_{[\sigma \beta]} h^{\alpha \sigma} \tau^{\nu \beta} \nabla_{\nu}^{\frak{C}} h_{\rho \alpha} + 2 \frak{C}^{\nu}_{[\sigma \beta]} \tau^{\gamma \beta} \frak{C}^{\sigma}_{[\gamma \nu]} \rp, \\
B^{b}_{\beta}\frac{\delta \bar{\LL}(0)}{\delta \tau^{b}_{\beta}} & = 0, \\
\psi_{\lambda \tau}\frac{\delta \bar{\LL}(0)}{\delta h_{\lambda \tau}} & = 0, \\
m_{\lambda}^{c}\frac{\delta}{\delta \tau^{c}_{\lambda}} \frac{\p \bar{\LL}(0)}{\p \sigma}\bigg|_{\sigma= 0} & = em_{\lambda}^{c} \Bigg[ \tau^{\lambda}_{c} h^{\sigma \nu} \frak{R}_{\sigma \nu} + \lp h^{\lambda \sigma} h^{\rho \nu} - h^{\lambda \rho} h^{\sigma \nu} \rp \lp 2 A^{b}_{c \rho} K_{\nu \sigma b} - 2 \nabla_{\rho}^{\frak{C}} K_{\nu \sigma c} - K_{\sigma \nu d} A^{b}_{e \rho} \eta^{ed}\eta_{cb} \rp  \nonumber \\
& \ph{=}  -4 \frak{C}^{\rho}_{[\rho \mu]} \frak{C}^{\beta}_{[\beta \alpha]} h^{\mu \alpha} \tau^{\lambda}_{c} + \lp h^{\lambda\mu} \tau^{\alpha}_{c} - \frac{1}{2}h^{\mu \alpha} \tau_{c}^{\lambda}\rp \lp 4 \frak{C}^{\beta}_{[\mu \nu]} \frak{C}^{\nu}_{[\alpha \beta]} + A^{a}_{d \mu} A^{b}_{e \alpha} \eta^{ed}\eta_{ab}\rp  \nonumber \\
&\ph{=} + \lp h^{\rho \alpha} \tau^{\lambda}_{d} - h^{\lambda \alpha} \tau_{d}^{\rho}\rp\lp 2 A^{d}_{c \alpha} \frak{C}^{\mu}_{[\mu \rho]} + 2 \frak{C}^{\beta}_{[\beta \rho]} A^{b}_{e \alpha} \eta^{ed} \eta_{cb} - \nabla_{\rho}^{\frak{C}}A^{b}_{e \alpha} \eta^{ed} \eta_{cb} \rp \nonumber \\
& \ph{=}+ 2 \tau^{\beta}_{c} \lp h^{\nu \alpha} \nabla_{\nu}^{\frak{C}} \frak{C}^{\lambda}_{[\alpha \beta]} - h^{\lambda \alpha} \nabla_{\nu}^{\frak{C}} \frak{C}^{\nu}_{[\alpha \beta]}\rp + h^{\mu \alpha} \tau^{\lambda}_{e} \lp A^{a}_{d \mu} \eta^{e d} \eta_{ab} A^{b}_{c \alpha} - A^{e}_{d \mu} A^{b}_{a \alpha} \eta^{ad} \eta_{cb}\rp  \nonumber \\
&\ph{=} +  4 \lp h^{\mu \alpha} \tau^{\lambda}_{c} - h^{\lambda \alpha} \tau^{\mu}_{c}\rp \nabla_{\mu}^{\frak{C}} \frak{C}^{\beta}_{[\beta \alpha]}\Bigg], \\
\Phi_{\sigma \lambda}\frac{\delta}{\delta h_{\sigma \lambda}} \frac{\p \bar{\LL}(0)}{\p \sigma} \bigg|_{\sigma=0} & = e \Phi_{\sigma \lambda}\Bigg[ \frac{1}{2} h^{\mu \nu} h^{\sigma \lambda} \frak{R}_{\mu \nu}- h^{\sigma \mu} h^{\lambda \nu} \frak{R}_{\mu \nu} + \lp 4 \frak{C}^{\alpha}_{[\alpha \nu]} \frak{C}^{\mu}_{[\mu \rho]} - 2 \nabla_{\nu}^{\frak{C}} \frak{C}^{\mu}_{[\mu \rho]}\rp \lp h^{\nu \sigma} h^{\rho \lambda} - h^{\lambda \sigma} h^{\rho \nu}\rp \nonumber \\
& \ph{=} + \frac{1}{4}\lp h^{\mu \alpha} h^{\sigma \lambda} - 2 h^{\mu \sigma} h^{\alpha \lambda}\rp\lp 8 \frak{C}^{\beta}_{[\beta \alpha]} \frak{C}^{\nu}_{[\nu \mu]} - 4 \frak{C}^{\beta}_{[\mu \nu]} \frak{C}^{\nu}_{[\alpha \beta]} - A^{a}_{d \mu} A^{b}_{c \alpha} \eta^{cd} \eta_{ab}\rp\Bigg], \\
\left[ \dots\right] & = - \frac{1}{4} e \eta_{cd} h^{\mu \rho} h^{\nu \sigma} F_{\mu \nu}^{c} F_{\rho  \sigma}^{d}.
\end{align}
\end{subequations}
By adding up all terms above and term with Lagrange multiplier which enforces (\ref{onshell}) 
\begin{equation}
\zeta_{\rho \sigma a} e h^{\mu \rho} h^{\nu \sigma} \p_{[\mu} \tau_{\nu]}^{a}
\end{equation}
we obtain final nonrelativistic Lagrangian, which concludes the main result of this paper.
\section*{Acknowledgements}
\noindent I would like to thank Josef Kluso\v{n} and Michal Pazderka for useful discussions and a lot comments to this paper. This work was supported from Operational Programme Research, Development and Education – „Project Internal Grant Agency of Masaryk University" (No.CZ.02.2.69/0.0/0.0/19\_073/0016943).

\end{document}